\documentclass[12pt]{article}
\usepackage{scicite}
\usepackage{times}
\usepackage{setspace} 
\usepackage{graphicx}
\newenvironment{sciabstract}{
\begin{quote} \bf}
{\end{quote}}

\makeatletter
\renewcommand*{\ps@plain}{
  \let\@mkboth\@gobbletwo
  \let\@oddhead\@empty
  \def\@oddfoot{
    \reset@font
    \hfil
    \thepage
  }
  \let\@evenhead\@empty
  \let\@evenfoot\@oddfoot
}
\makeatother
\title{A Class of Warm Jupiters with Mutually Inclined, Apsidally Misaligned, Close Friends} 
\author
{Rebekah I. Dawson,$^{1\ast}$ Eugene Chiang$^{1}$\\
\\
\normalsize{$^{1}$Department of Astronomy, University of California, Berkeley,}\\
\normalsize{501 Campbell Hall \#3411, Berkeley, CA 94720-3411, USA}\\
\\
\normalsize{$^\ast$To whom correspondence should be addressed; E-mail:  rdawson@berkeley.edu.}
}

\date{}

\begin{document}

\baselineskip24pt

\maketitle 

\begin{sciabstract}
  The orbits of giant extrasolar planets often have surprisingly small
  semi-major axes, large eccentricities, or severe misalignments
  between their normals and their host stars' spin axes. In some
  formation scenarios invoking Kozai-Lidov oscillations, an external
  planetary companion drives a planet onto an orbit having these
  properties. The mutual inclinations for Kozai-Lidov oscillations can
  be large and have not been confirmed observationally. Here we deduce
  that observed eccentric warm Jupiters with eccentric giant
  companions have mutual inclinations that oscillate between
  35--65$^\circ$. Our inference is based on the pairs' observed apsidal
  separations, which cluster near 90$^\circ$. The near-orthogonality
  of periapse directions is effected by the 
outer companion's
quadrupolar
  and octupolar potentials. These systems may be undergoing a stalled
  version of tidal migration that produces warm Jupiters over hot
  Jupiters, and provide evidence for a population of multi-planet
  systems that are not flat and have been sculpted by Kozai-Lidov
  oscillations.
\end{sciabstract}

Planet-planet scattering \cite{ras96}, secular chaos \cite{wu11}, and
Kozai-Lidov oscillations \cite{koz62,lid62} induced by a stellar
\cite{wu03} or planetary \cite{nao11} perturber are each capable of
exciting the eccentricities of giant planets 
\cite{tak05,jur08}, of shrinking orbits by tidal friction to
form close-in hot Jupiters (having semimajor axes $a < 0.1$ AU), and
of tilting hot Jupiters' orbit normals away from their host stars'
spin axes \cite{fab07,wu07,cha08,nag11,mor11,nao12}.

There are two outstanding issues with these models. First, they
require or produce large inclinations between planetary orbits. These
have not yet been observed.  Most of the few systems with measured
mutual inclinations are composed of planets on co-planar, low
eccentricity, mean-motion resonant orbits, e.g., GJ 876 \cite{riv10} and 
Kepler-30 \cite{san12}. This breed of system is thought to result from
gentle disk migration \cite{lee02}, not from Kozai oscillations,
planet scattering, or secular chaos. Two notable exceptions contain
well-spaced (i.e., non-resonant) giant planets. Based on
astrometry with the Hubble Space Telescope fine guidance sensor, a
mutual inclination of $30^\circ$ between Upsilon Andromeda c and d was
inferred \cite{mca10}. 
From transit timing variations, a mutual inclination 
of 9$^\circ~^{+8}_{-6}$ between 
Kepler-419 (KOI-1474) b and c
was measured \cite{daw14}.

The second problem is that current models do not easily produce warm
Jupiters, located exterior to hot Jupiters but interior to the pile-up
of giant planets at 1 AU \cite{cum08}. 
Although intrinsically rare,
warm Jupiters
promise
to distinguish among models
by serving as the exception that proves the rule. Many warm Jupiters
have present-day eccentricities too high to have resulted from
planet-planet scattering because giant planets at small orbital
distances collide and circularize before their velocity dispersions
become too elevated \cite{pet14a}. Yet most of their observed
eccentricities are also too low to be 
easily accommodated
within formation
scenarios for hot Jupiters that invoke tidal migration \cite{soc12},
as warm Jupiters' current periapses are too far from their host stars
for tidal friction to operate effectively.

One possibility \cite{don14} is that eccentric warm Jupiters are
undergoing
``slow Kozai'' \cite{pet14} migration driven by an
external, mutually inclined companion. In this interpretation, warm
Jupiters are observed today at the low-eccentricity phases of their
secular Kozai cycles, and only rarely attain eccentricities high
enough for tidal friction to operate. For Kozai oscillations to not be
quenched at these small semi-major axes by general relativistic
precession, the external perturber must be nearby: it must be a
``close friend'' \cite{don14}, in contrast to a ``cold friend''
\cite{knu14}. Supporting the possibility of migration driven by close
friends, eccentric warm Jupiters orbit stars more enriched in metals
--- and therefore more likely to host multiple giant planets --- than
those of their circular counterparts \cite{daw13}. Indeed those warm
Jupiters known to have external giant planetary companions exhibit a
broader range of eccentricities than solitary warm Jupiters
\cite{don14}. A key question is whether the mutual orbital inclination
$i_{\rm mut}$ between warm Jupiters and their exterior companions
exceeds 39.2$^\circ$, the minimum angle required to excite Kozai
oscillations in the quadrupole approximation
\cite{koz62,lid62}. Unfortunately, nearly all these planets are
detected by the radial velocity (RV) method, which does not yield
$i_{\rm mut}$ directly, but instead measures a given planet's $a$,
eccentricity $e$, minimum mass $m\sin i_{\rm sky}$ (where $i_{\rm
  sky}$ is the orbital inclination with respect to the sky plane), and
argument of periapse $\omega_{\rm sky}$ (referred to the sky plane).

However, the sky-projected apsidal separation of a planetary pair,
$\Delta \omega_{\rm sky}$, can be a clue to $i_{\rm mut}$
\cite{chi01}. In the invariable plane, the difference in apsidal
longitudes, $\Delta \varpi_{\rm inv}$ (for which $\Delta \omega_{\rm
  sky}$ is our observable proxy), is often found near 0$^\circ$ or 180$^\circ$
for pairs of co-planar planets, either in the secular limit
\cite{mic04} or in the 2:1 mean-motion resonance \cite{lee02}. By
contrast, for highly inclined systems, there is no such preference to
find $\Delta \varpi_{\rm inv}$ near 0$^\circ$ or 180$^\circ$, at least for
secular systems \cite{chi01}.
As will be shown, the behavior of $\Delta \varpi_{\rm inv}$ is
directly reflected by its projection $\Delta \omega_{\rm sky}$. Fig.~1a
displays $|\Delta \omega_{\rm sky}|$ for RV-detected planetary pairs
\cite{eod} --- some of which include warm Jupiters but most of which
do not --- with well-constrained \cite{unc} $e$ and $\omega_{\rm
  sky}$, as a function of angular momentum ratio. Most systems with
angular momentum ratio $>$ $\sim$3 have $|\Delta \omega_{\rm sky}|$ near
0$^\circ$ and 180$^\circ$. But some pairs with similarly large ratios are
clustered near 90$^\circ$.

Intriguingly, the cluster of systems having $|\Delta \omega_{\rm sky}|
\sim 90^\circ$ includes eccentric warm Jupiters with eccentric close
friends. In fact, if we turn the problem around and consider only
those systems with eccentric pairs consisting of one warm (0.1 $< a <
1$ AU) and one ``balmy'' ($a > 1$ AU) Jupiter (defined as $m \sin
i_{\rm sky} > 0.1 M_{\rm Jupiter})$ without regard for $|\Delta
\omega_{\rm sky}|$, then of the eight systems \cite{select} so
selected (red symbols in Fig.~1a), six (red
circles)
have $|\Delta \omega_{\rm sky}|$
near 90$^\circ$ (HD 147018, HD 38529, HD 168443, HD 74156, HD 169830,
and HD 202206).
The two systems we know or expect to have low mutual inclinations (red crosses) are not
among
these
six pairs with $|\Delta \omega_{\rm sky}|$ clustered near 90$^\circ$.
HD 82943 (bottom red cross), is librating in the 2:1 mean motion resonance \cite{tan13},
which interferes with the purely secular interactions studied here by
driving
apsidal precession on a shorter, resonant timescale. Another
outlier is the transit-detected system  Kepler-419 (top red cross)
which was recently found to host an eccentric pair of one warm and one balmy
Jupiter having a low ($9^\circ~^{+8}_{-6}$) mutual inclination
and an apsidal separation of $|\Delta \omega_{\rm sky}|=179^\circ.8^{+0.6}_{-0.7}$ \cite{daw14}.
The orbital elements of all eight systems are listed in Table S1.

Here we argue that the $\sim$90$^\circ$ apsidal misalignment between
warm Jupiters and their close friends signifies a mutual inclination
of $\sim$40$^\circ$, just above the lower limit for Kozai
oscillations. For each of the six systems identified above\cite{hd206}, we perform $\sim$1000 numerical orbit integrations,
starting with initial conditions that randomly assign the two angles
not constrained by the radial-velocity data, $i_{\rm sky}$ and
$\Omega_{\rm sky}$, the latter being the longitude of ascending node
on the sky plane (with the mass $m$ chosen to satisfy the measured
$m\sin i_{\rm sky}$). These angles are drawn (independently for each
planet) from a uniform distribution spanning $-1<\cos i_{\rm sky}<1$
and $0^\circ < \Omega_{\rm sky} < 360^\circ$, resulting in an isotropic
distribution of orbits.
For an additional 180 simulations, we fix $i_{\rm sky,1} = i_{\rm
  sky,2} = 90^\circ$
and $\Omega_{\rm sky,2}=0^\circ$, and draw
$\Omega_{\rm sky,1}$ from a uniform distribution spanning $0^\circ$ to
180$^\circ$. The eccentricities, semi-major axes, mean anomalies, and $\omega_{\rm sky}$'s are set to those observed. All numerical
integrations in this paper use the $N$-body code {\tt Mercury6}
\cite{cha99} with the Bulirsch-Stoer integrator ({\tt bs}), modified
to include general-relativistic precession \cite{fab11}.

In Fig.~1b, we plot the fraction of time each
simulated stable system spends having $75^\circ < |\Delta \omega_{\rm sky}| < 135^\circ$, as a function of
median mutual inclination,
for five of our six systems (the sixth is a special case discussed below).
Note that $\Delta \omega_{\rm sky}
=270^\circ=-90^\circ$ is equivalent to 90$^\circ$ in that both yield
$|\Delta \omega_{\rm sky}|$ = 90$^\circ$.  Each simulation is
inspected by eye to ensure that the integration duration is long
enough to sample the behavior of $\Delta \omega_{\rm sky}$; integration durations
range 
from 0.15 to 100 Myr. A similar exploration of parameter space \cite{vf10}
was performed for two of our systems, HD 38529 and HD 168443, but without
our aim of identifying regions of parameter space where
$|\Delta \omega_{\rm sky}|$ lingers near $90^\circ$.  
As illustrated in Fig.~1b,
in 
the five
systems, we encounter
configurations in which $|\Delta \omega_{\rm sky}|$ spends excess time
near 90$^\circ$ --- but only for median
$i_{\rm mut} \sim 39^\circ$, near Kozai's critical inclination.
For such systems,
$|\Delta \omega_{\rm sky}|$ spends 
1/2 -- 2/3 of the time
between 75$^\circ$ and 135$^\circ$, depending on initial conditions. By
comparison, uniform circulation of $|\Delta \omega_{\rm sky}|$ gives a
fractional time of 1/3. Nearly co-planar systems for which $|\Delta
\omega_{\rm sky}|$ librates about 0$^\circ$ or 180$^\circ$ may spend
no time at all
in
the desired range. Such
realizations
are inconsistent with
the radial-velocity data and accordingly do not appear in
Fig.~1b.  
The sixth system, HD 202206,
is distinctive because its warm Jupiter
is much more massive than the
outer companion and because it lies near the 5:1 resonance.
For this system, we still
encounter configurations with $i_{\rm mut} \sim 40^\circ$
spend that excess time
with $|\Delta \omega_{\rm sky}|$ near 90$^\circ$, but
not in the conventional range of 75--135$^\circ$.
An example shown in Fig.~S5.

We reiterate that those integrations exhibiting libration of $|\Delta
\omega_{\rm sky}|$ about 90$^\circ$ all have $i_{\rm mut}$ close to
39.2$^\circ$; typically $i_{\rm mut}$ varies within an interval that does not fall outside 35--65$^\circ$ over
the course of a given integration. 
In contrast, 
nearly co-planar, polar, or retrograde
configurations consistent with the radial-velocity data have
$|\Delta \omega_{\rm sky}|$ circulating in the integrations we
performed. Fig.~2 shows examples of how the apsidal separation lingers near
$90^\circ$ and $270^\circ$, in the sky plane and invariable plane,
over many 
secular
oscillations.  
The desired libration of $|\Delta
\omega_{\rm sky}|$ does not require a finely-tuned viewing geometry or
set of initial conditions.
Regarding viewing geometry: the
libration of
$|\Delta \omega_{\rm sky}|$ is similar to that of $|\Delta \varpi_{\rm inv}|$
for most observer orientations.
Taking a single integration of HD 147018 for which
$|\Delta \varpi_{\rm inv}|$
spends 60$\%$ of its time between 75$^\circ$
and 135$^\circ$ in the invariable plane,
we
viewed this system from the vantage points
of 500 isotropically distributed observers.
We find that
60\% (80\%) of observers
measure
$75^\circ < |\Delta \omega_{\rm sky}| < 135^\circ$
at least 50\% (40\%) of the time.
Measurements from four example observers are given in the right
column of Fig.~2.
Regarding initial conditions:
we numerically integrate
different realizations
of the five systems shown in Fig.~2,
ignoring all
observational constraints on $\omega_{\rm sky}$ and starting from
various combinations of initial $\{\varpi_{\rm inv,1}, \varpi_{\rm
  inv,2}, i_{\rm mut}\}$, with 
$35^\circ < i_{\rm mut} < 65^\circ$. To simplify and speed up these calculations,
we approximate the inner planet as a test particle
and integrate the secular equations of motion (described further below).
Among the many realizations thus constructed (three are shown in Fig. S6 along with accompanying surfaces of section),
we find that it is not unusual for $\Delta \varpi_{\rm inv}$ (and by extension
$\Delta \omega_{\rm sky}$) to linger near 90$^\circ$/270$^\circ$ ---
i.e., it is not uncommon for $\Delta \varpi_{\rm inv}$ to librate
about 180$^\circ$ with a libration amplitude of 90$^\circ$. Actually,
depending on initial conditions, $\Delta \varpi_{\rm inv}$ can librate
about either $180^\circ$ or $0^\circ$ with a variety of amplitudes, as
well as circulate, and we discuss later why warm Jupiters may prefer a
libration amplitude near 90$^\circ$.

The apsidal libration observed here has been seen in other celestial
mechanical contexts; it is qualitatively similar to the
``artichoke-shaped'' \cite{whi93} libration exhibited by Saturn's
irregular satellites Narvi \cite{cor10} and, for some initial
conditions, Pasiphae \cite{whi93,yok03}. Both Narvi and Pasiphae
orbit Saturn with inclinations of $\sim$140$^\circ$ with respect to
the plane of Saturn's orbit about the Sun. In the context of triple
stellar systems, ``beatlike'' patterns
with superposed short and long-timescale oscillations,
similar to those observed in Fig.~2,
were noticed in the modeled eccentricity
variations
of systems with
$i_{\rm mut}$ $\sim$ 40$^\circ$, and attributed to interference
between the quadrupolar and octupolar potentials \cite{for00}. 
The configurations of interest here do not lie near any border between
circulation and libration; i.e., $\Delta \varpi_{\rm inv}$ librates
about 180$^\circ$ with an amplitude of 90$^\circ$, not $180^\circ$. In
particular, the dynamics responsible for how $|\Delta \varpi_{\rm inv}|$
lingers near $90^\circ$ is dissimilar from the ``borderline'' behavior
of nearly coplanar systems in which $|\Delta \varpi_{\rm inv}|$ speeds
through $180^\circ$ but does not linger near 90$^\circ$
\cite{bar07,for08}.

We can reproduce the observed libration of $\Delta \varpi_{\rm inv}$
by approximating a warm Jupiter as a test particle perturbed by an
exterior body, and solving Lagrange's equations of motion using a
secular
disturbing potential expanded to octupolar order \cite{yok03},
including general relativistic precession of both planets' orbits
\cite{fab11}.  The octupolar potential plays an essential role in
generating a precession rate d$\Delta \varpi_{\rm inv}$/dt that
cancels, on average, the precession rate induced by the quadrupolar
potential; the net result is that $\Delta \varpi_{\rm inv}$
librates. Therefore the perturber must not only be near enough to
dominate general relativistic precession \cite{wu03,don14} but must
also be eccentric: the strength of the octupolar potential is
proportional to the perturber's eccentricity.  The importance of the
octupolar potential has only recently been recognized in the exoplanet
literature \cite{for00,nao11}.  A popular application has been to flip
an interior body from a prograde to retrograde orbit at large
eccentricities \cite{kat11,lit11,nao11,nao13,li14,li14b}. A low mutual
inclination, well below
Kozai's
critical angle, is sufficient to spur such a flip if the inner
planet's
eccentricity is high enough \cite{li14,li14b}. The regime explored here
--- which produces warm Jupiters with the observed clustering
of $|\Delta \omega_{\rm sky}|$ near $90^\circ$ --- is complementary:
both the inner and outer planets' eccentricities are too modest
for the octupolar potential to effect flips,
and $i_{\rm mut}$ remains near 40$^\circ$,
a prograde configuration. Indeed we deduce that our six systems with warm Jupiters and close friends are prograde with $i_{\rm mut}$ $\sim$ 40$^\circ$ rather than retrograde with $i_{\rm mut}$ $\sim$ 140$^\circ$, because in the latter case
$\omega_{\rm sky,1} + \omega_{\rm sky,2}$ would librate rather than
$\omega_{\rm sky,1} - \omega_{\rm sky,2}$ \cite{yok03}.

It is no coincidence that our six warm Jupiter systems are
all characterized by mutual inclinations abutting Kozai's minimum
angle. In our regime in which the quadrupole potential still dominates
and planetary eccentricities are
low enough to avoid flips,
$39^\circ$ is the inclination that coincides with maximum eccentricity 
(minimum periapse) 
and hence maximum tidal dissipation. Orbital decay during this
maximum-eccentricity, minimum-inclination phase of the Kozai cycle
naturally leads to an abundance of tidally migrated warm Jupiters with
$i_{\rm mut}$ near 
39--65$^\circ$ \cite{kis98}. The octupolar potential
is not strong enough
for our systems
to alter this feature of quadrupolar Kozai cycles
\cite{nao11,kis98}. 
At the same time,
the reason why warm Jupiters have not
completed their migration to become hot Jupiters is because of the
special octupole-modified nature of their eccentricity variations. The
usual (quadrupole) Kozai oscillations of eccentricity, which occur
over the short nodal precession period, are modulated by an
octupole-induced
envelope\cite{hd206env} of much longer period following that of
apsidal libration (Fig.~3; Figs.~S1-S5). The envelope period is
approximately $(a_2/a_1) (1-e_2^2)/e_2$ longer than the Kozai
timescale \cite{li14}.  This long-period modulation prevents
eccentricities from surging too often, and renders migration even
slower than the ``slow Kozai'' migration described in  \cite{pet14}. Such ``super-slow''
evolution is similar to the ``step'' migration seen at high $i_{\rm
  mut}$ \cite{nao11}, except without the transition from prograde to
retrograde orbital motion and the accompanying rapid tidal
circularization. In the gentle and intermittent migration considered
here, warm Jupiters reach the small periapses characterizing hot
Jupiters only at the peaks of their eccentricity envelopes.  As a
proof of concept, we perform an integration of the 
secular
equations of motion
including tidal friction (Fig.~4).  
The warm Jupiter undergoes
super-slow tidal migration in which it librates apsidally and stalls
in semi-major axis. 
A libration amplitude
for $\Delta \varpi_{\rm inv}$ 
near 90$^\circ$ enables
super-slow migration.
If the libration amplitude were smaller, or if the apsidal separation were to circulate, then the envelope modulating the eccentricity would be less peaky, and the interior planet would spend
more of its time near its maximum eccentricity.
If the libration amplitude were larger, 
then not only would the apsidal separation be more prone to circulate given
a small perturbation, but 
the system would 
also 
be more vulnerable to
retrograde flips and concomitant eccentricity surges ($e\rightarrow1$).
The upshot
of all these scenarios in which $\Delta \varpi_{\rm inv}$ does not
librate with an amplitude of $90^\circ$
is that tidal migration, once begun, would rapidly complete
and spawn a hot Jupiter (see Fig.~S7).

The class of warm Jupiters we have identified is similar to the
predicted class of Kozai-oscillating warm Jupiters \cite{don14},
except that here the octupolar field generated by the eccentricity of
the exterior perturber plays a starring role in effecting the
observed near-orthogonality of periapse directions and in braking
tidal migration. The mutual inclination of $i_{\rm mut}$ $\sim$
35--65$^\circ$ we have inferred from the measured apsidal misalignment
attests to how the Kozai mechanism,
working between planets, has indeed shaped planetary systems. Although
large inclinations between hot Jupiters' orbital planes and the
equatorial planes of their host stars are well established, our
finding provides evidence that pairs of more distant giant planets are
themselves highly mutually inclined (although still prograde), in
stark contrast to the flatness of solar system planets. The origin of
such large inclinations is a mystery; planet-planet scattering \cite{ras96} or
secular chaos are possibilities \cite{wu11}.

\clearpage
\begin{centering}
\noindent\includegraphics{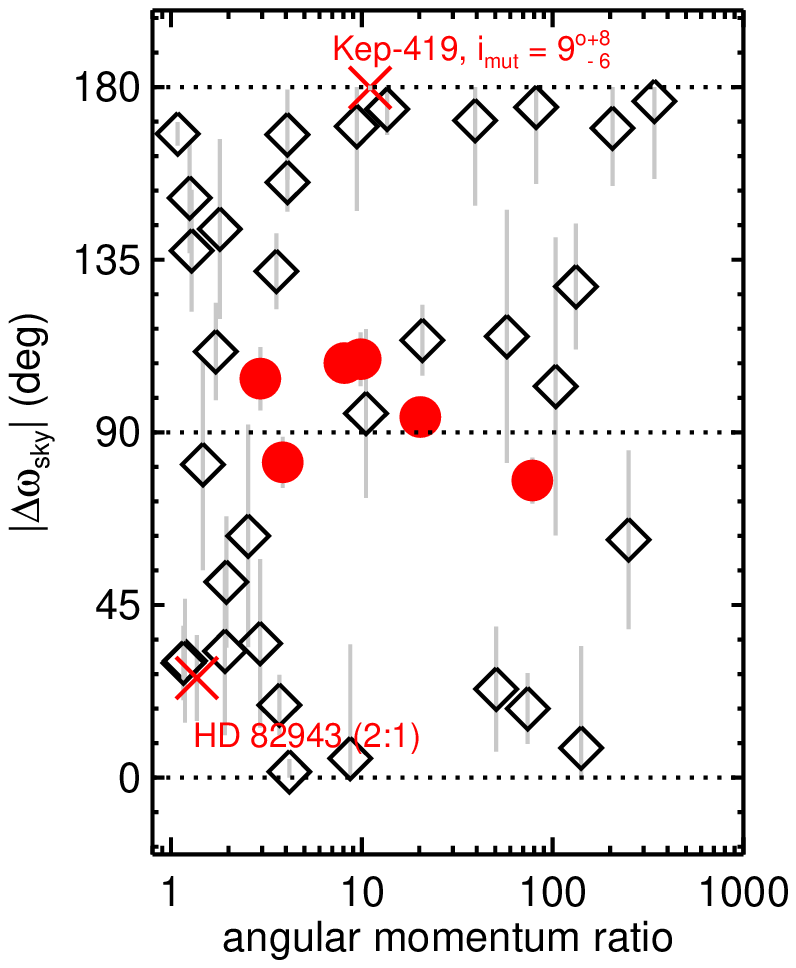}\includegraphics{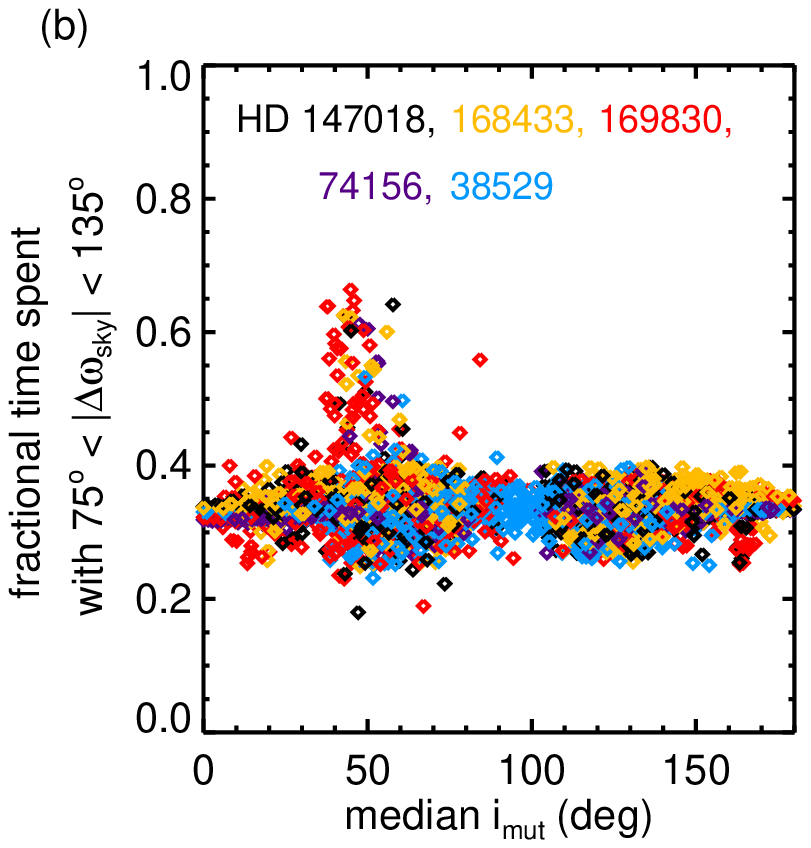}\\
\end{centering}
\clearpage {\noindent {\bf Fig.~1.} 
Apsidal misalignments between pairs of planets, and the large mutual
inclinations implicated.
Left (panel a): Sky-projected periapse
  separations $\Delta \omega_{\rm sky}$
of planetary pairs,
where ${-180^\circ<\Delta\omega_{\rm sky}<180^\circ}$. 
 The sample includes all RV- and
  transit-discovered pairs with well-constrained\cite{unc}
  eccentricities and arguments of periapse.
No mass or semi-major axis cut is made on the black
diamonds.
The 
red circles
(corresponding to systems shown also in Figs.~2--3 and S1--S5) and
crosses
(HD 82943, Kepler-419) represent warm Jupiters with one and
  only one known companion beyond 1 AU. All six of the red 
circles\cite{others}
  lie close to $|\Delta \omega_{\rm sky}| = 90^\circ$, 
i.e., their apsides are strongly misaligned.
For calculation
  of the abscissa values, the orbital angular momentum is evaluated as
  $m \sin i_{\rm sky} \sqrt{a(1-e^2)}$. 
Right (panel b): Fraction of time 
that $|\Delta \omega_{\rm sky}|$ spends near 90$^\circ$,
as a function of median mutual inclination,
for stable
simulations of systems corresponding to five of the six red circles
(the same five are shown in Fig.~2).
HD 202206 is omitted; its architecture and apsidal
behavior differ from those of the five (see Fig.~S5).
Each simulation is
  consistent with the observed orbital elements, but only those with
  $i_{\rm mut}$ near 40$^\circ$ spend extra time at
  $|\Delta\omega_{\rm sky}| \sim 90^\circ$.
The fractional time is evaluated
for $75^\circ<|\Delta\omega_{\rm sky}|<135^\circ$;
this window is centered on 105$^\circ$ instead of 90$^\circ$
because of general relativistic precession.}  \clearpage

\includegraphics[width=6in]{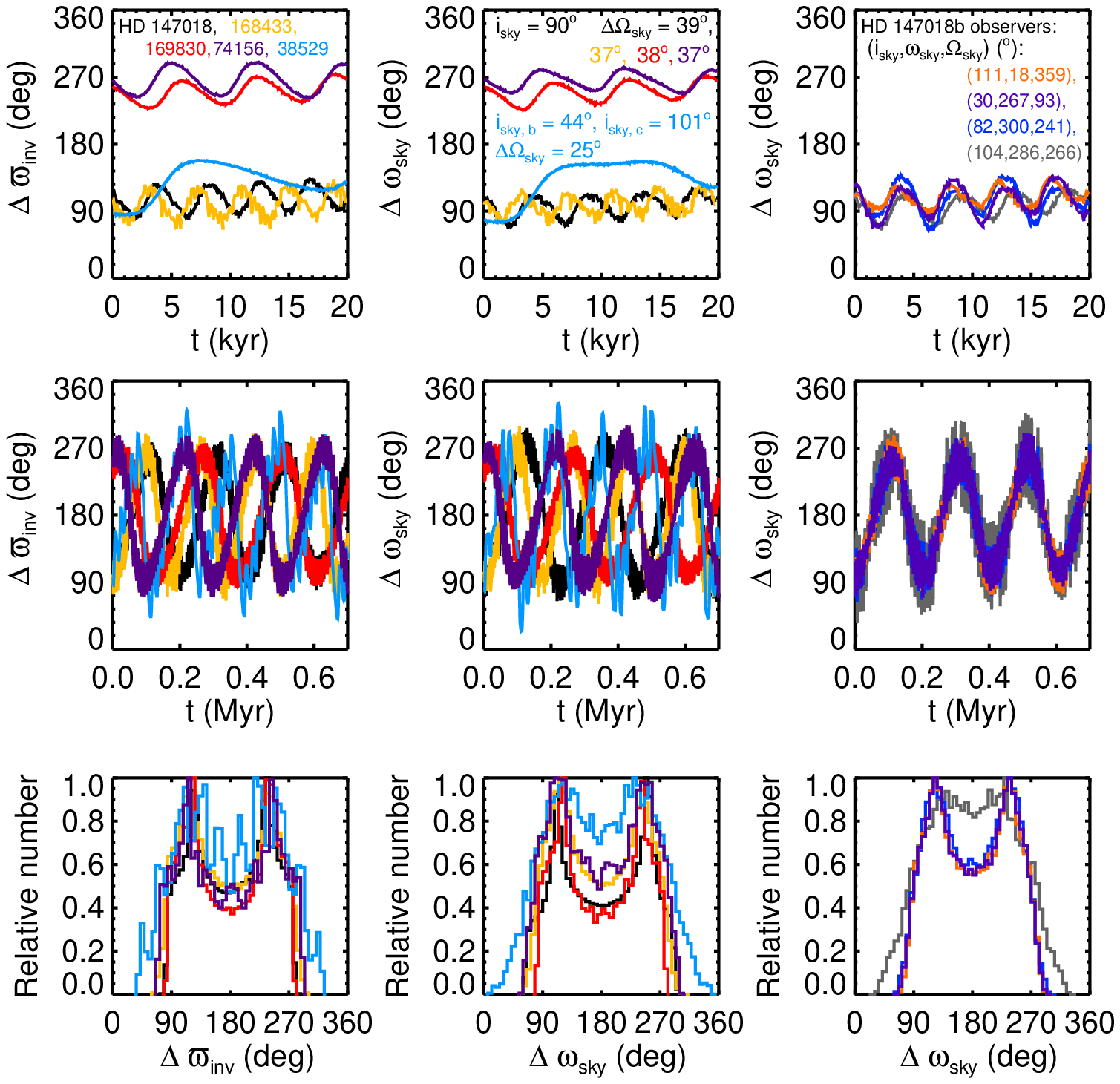}\\
{\noindent {\bf Fig.~2.} Sample time histories of apsidal
misalignment for five planetary pairs of warm Jupiters and their outer companions.
Initial conditions are
taken from Table S1 with additional orbital angles indicated in the legend.
For every history shown, the mutual inclination between the pair of planets varies within an interval that does not fall outside of 35--65$^\circ$.
 Top: On short timescales, the apsidal
separation lingers
near
90$^\circ$ or 270$^\circ$. Middle: Over
longer timescales, the apsidal misalignment librates about
180$^\circ$
with an amplitude of 90$^\circ$.
Bottom: Relative occurrence of
either the apsidal separation evaluated in the invariable plane
($\Delta \varpi_{\rm inv}$) or the
sky-projected apsidal
separation ($\Delta \omega_{\rm sky}$), in uniformly spaced bins,
indicating that
these angles
spend
excess time near
90$^\circ$ and 270$^\circ$.  Because of the lower oscillation
frequencies characterizing HD 38529 and HD 74156, 
for these two systems
we plot 
$t/3$
as the
abscissa
in both top and middle rows.
Unlike the systems shown here,
HD 202206 is situated near the 5:1 resonance and its
inner planet is almost certainly more massive than its outer planet (see Table S1);
its dynamics is explored in Fig.~S5.
}

\clearpage

\clearpage
\includegraphics[width=6in]{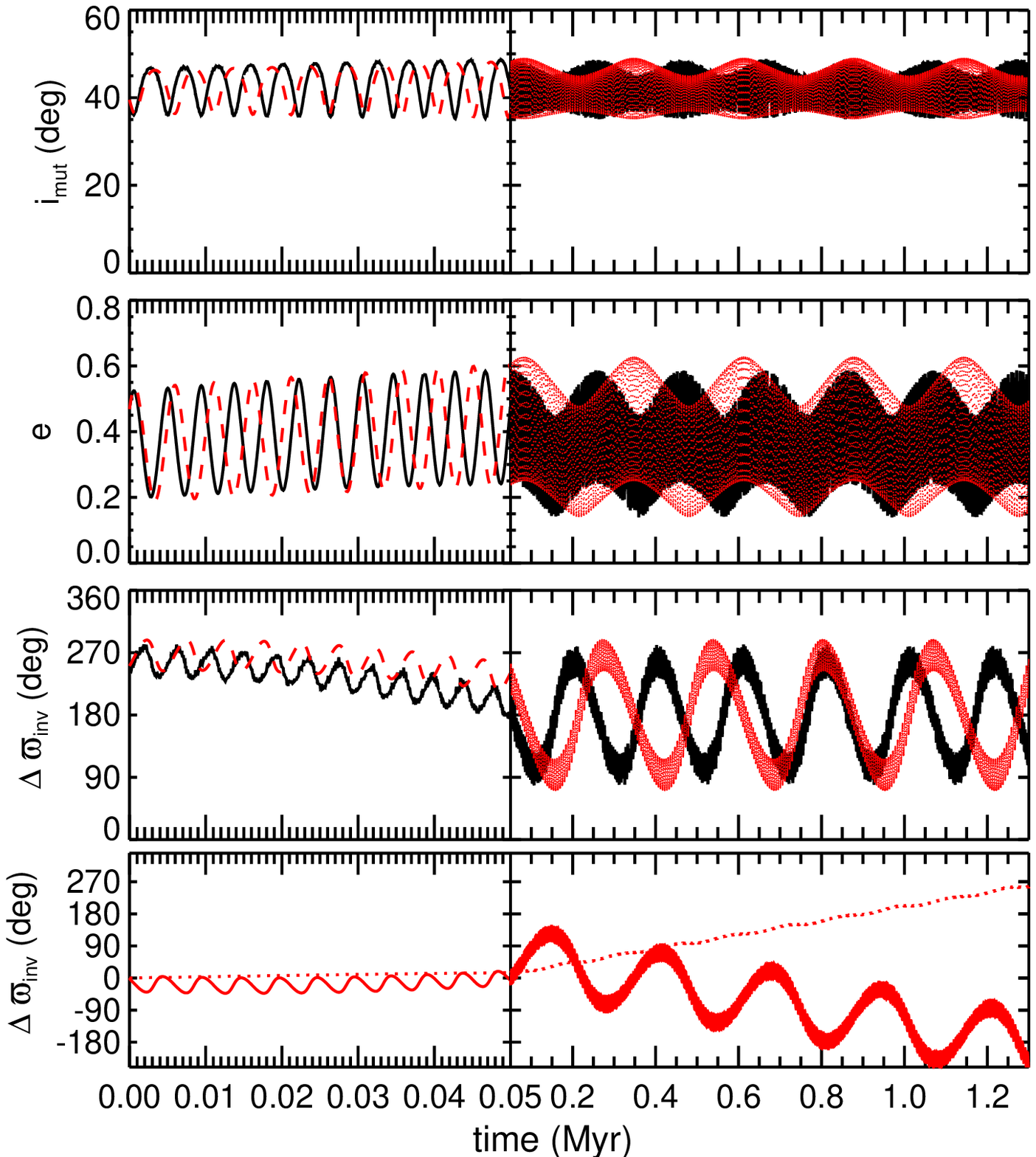}\\
{\noindent {\bf Fig.~3.} 
Orbital evolution of HD 147018b, calculated using an $N$-body simulation and using a secular Hamiltonian expanded to octupolar order.
Plotted are
  the time evolution of the mutual inclination (row 1), eccentricity
  of the inner warm Jupiter (row 2), apsidal separation (row 3), and
  near-canceling contributions to the apsidal separation from the
  quadrupolar and octupolar 
potentials
(row 4). The simulation is performed in the invariable plane.
 Initial conditions are for
  the HD 147018 system (Table S1) with $i_{\rm sky} = 90^\circ$
and
$\Delta \Omega_{\rm sky} = 39^\circ$, corresponding to $i_{\rm mut}$ = 39$^\circ$
and to angles
$\{i_1=35.6^\circ, \omega_1=66.0^\circ, \Omega_1=0^\circ\}$
and $\{i_2=3.4^\circ, \omega_2=136.9^\circ, \Omega_2=180^\circ\}$
in the invariable plane. All black
  curves in this figure are taken from the same {\tt Mercury6}
  $N$-body integration underlying the 
black curves in the left column of
  Fig.~2. Red curves are for a test particle integrated using
  Lagrange's equations of motion for the octupolar Hamiltonian
  \cite{yok03}. The bottom row was computed by separating the
  Hamiltonian into the quadrupolar terms (dotted line) and octupolar
  plus general-relativistic terms (solid line); by evaluating their
  respective contributions to $\dot{\varpi}_1$ using
  Lagrange's equations of motion;
  and by integrating
  $\dot{\varpi}_1$, setting the constant of integration to zero
  for simplicity. (The quantities $e$ and $i_{\rm mut}$ were
  computed from the full Hamiltonian.) The sum of the dotted and solid
  curves in the bottom row equals the deviation of the red curve in the
  third row from its initial value. 
The secular behavior of the test particle 
reproduces qualitatively well that of the full $N$-body integration.}
\clearpage

\includegraphics[width=6in]{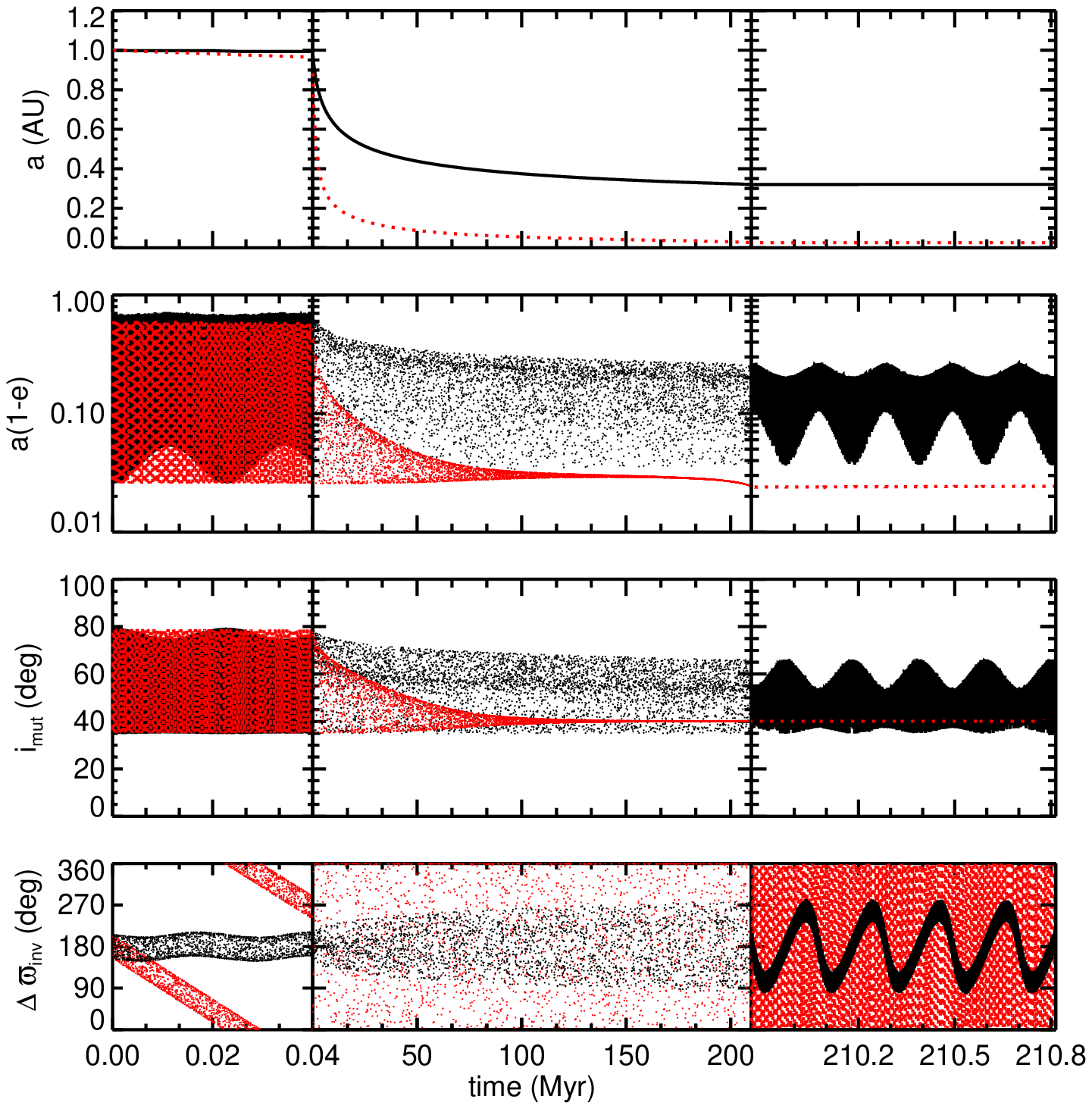}\\
{\noindent {\bf Fig.~4.} Tidal migration 
under a secular potential expanded beyond quadrupolar order
can
produce a stalled warm Jupiter. 
Plotted are the
time evolution of the inner planet's semi-major axis,
  periapse distance, mutual inclination with the outer planet, and
  apsidal separation, integrated using the test particle Hamiltonian
expanded
  to hexadecapolar order \cite{yok03}; 
practically identical results are obtained with the octupolar Hamiltonian.
Overplotted are results from the
  Hamiltonian including only up to quadrupolar terms (red
  dashed)
in which the inner planet fails to stall and instead becomes a hot Jupiter.
 General relativistic precession is included for both
  planets \cite{fab11}. Tidal evolution is implemented using a
  constant tidal quality factor of $10^5$, a Love number of 0.26, and
  a planetary radius of 1 Jupiter radius \cite{wu03}. Tides raised on
  the star are not included because they are weak compared to tides
  raised on the planet at
these semi-major axes.
The tidal forcing
  frequency is set to $\sqrt{\frac{G
      (M_\star+m_1)}{[a_1(1-e_1^2)]^3}}$ \cite{soc12}. The outer
  planet has $a_2= 1.923$ AU, $e_2= 0.133$, and $m_2= 6.59 M_{\rm Jupiter}$,
matching HD 147018c, and initial
$\{i_2=0^\circ, \omega_2=17.2^\circ, \Omega_2=180^\circ\}$. 
The inner planet (test particle) has initial $a_1=1$ AU,
$e_1=0.9$, and $\{i_1=65^\circ, \omega_1 = 38.4^\circ, \Omega_1=0^\circ\}$.  
With these choices, the eccentricity of the inner planet reaches
a minimum of 0.33 at $t=220$ yr during the first Kozai cycle.
The early tidal
  evolution, over the first $\sim$20 Myr, is subject to planet-planet
  scattering in a full $N$-body treatment; as such, the origin story
  portrayed in this figure is meant only to illustrate the concept of
  stalling, not to be definitive. This figure ends at $\sim$200 Myr
  but similar histories spanning a few Gyr are just as possible for
different initial conditions or tidal efficiency factors.}

\bibliographystyle{Science}

\begin{thebibliography}{10}
\bibitem{select}
We select for eccentric warm Jupiters with only one known companion, as
  additional companions can disrupt Kozai oscillations. Our selection criterion
  thus excludes planetary pairs that are members of the 55 Cnc, upsilon
  Andromeda, mu Arae, HIP 14810, Kepler-9, and Kepler-30 systems. Moreover,
  each of these systems also has a close-in ($a < 0.1$ AU) planet that would be
  disrupted if the warm Jupiter were to attain periapse distances small enough
  for tidal circularization.
\bibitem{hd206env}
The eccentricity of HD 202206b does not have such an envelope (Fig. S5), which
  may be due to its proximity to the 5:1 mean-motion resonance and the large
  relative mass of the inner planet.
\bibitem{others}
Among the black diamonds, the planetary pairs with $|\Delta\omega_{\rm sky}|$
  near $90^\circ$ and large angular momentum ratios include giant planets in a
  five-planet system (55 Cnc), a pair of hot super-Earths (GJ 163), and a cold
  Jupiter partnered with a hot Neptune (HD 125612); it is unclear whether their
  orthogonal eccentricity vectors also signify large mutual inclinations.
  The
transiting pair Kepler-30 c and d, which are near a 2:1 resonance, have
  $|\Delta\omega_{\rm sky}| = 114^\circ$,
an angular momentum ratio of 21, and a low
  mutual inclination \cite{san12}.
\bibitem{corrected}
We use the corrected elements reported by the EOD.
\bibitem{best}
Best-fit $N$-body solution \cite{cou10}, for which we have adjusted $a_2$ to
  obtain the results shown in Fig.~S5.
\bibitem{third}
In an independent analysis of the data, evidence was found for a third planet,
  not listed here \cite{bal14}.
\end{thebibliography}

{\bf \noindent Acknowledgements:} We gratefully acknowledge support from the Miller Institute for Basic Research in Science, the University of California (UC) Berkeley's Center for Integrative Planetary Science, the National Science Foundation, and the National Aeronautics and Space Administration. This research has made use of the Exoplanet Orbit Database at exoplanets.org. This research employed the SAVIO computational cluster provided by the Berkeley Research Computing program, which is supported by UC Berkeley's Chancellor, Vice Chancellor for Research, and Chief Information Officer. We thank four anonymous referees for  constructive feedback and Daniel Fabrycky and John Johnson for helpful comments.

\section*{Supplementary Materials}

\noindent Figures S1--S7

\noindent Table S1

\noindent References (52--63)
\clearpage

 \singlespacing 
\setcounter{page}{2}
\includegraphics[width=6in]{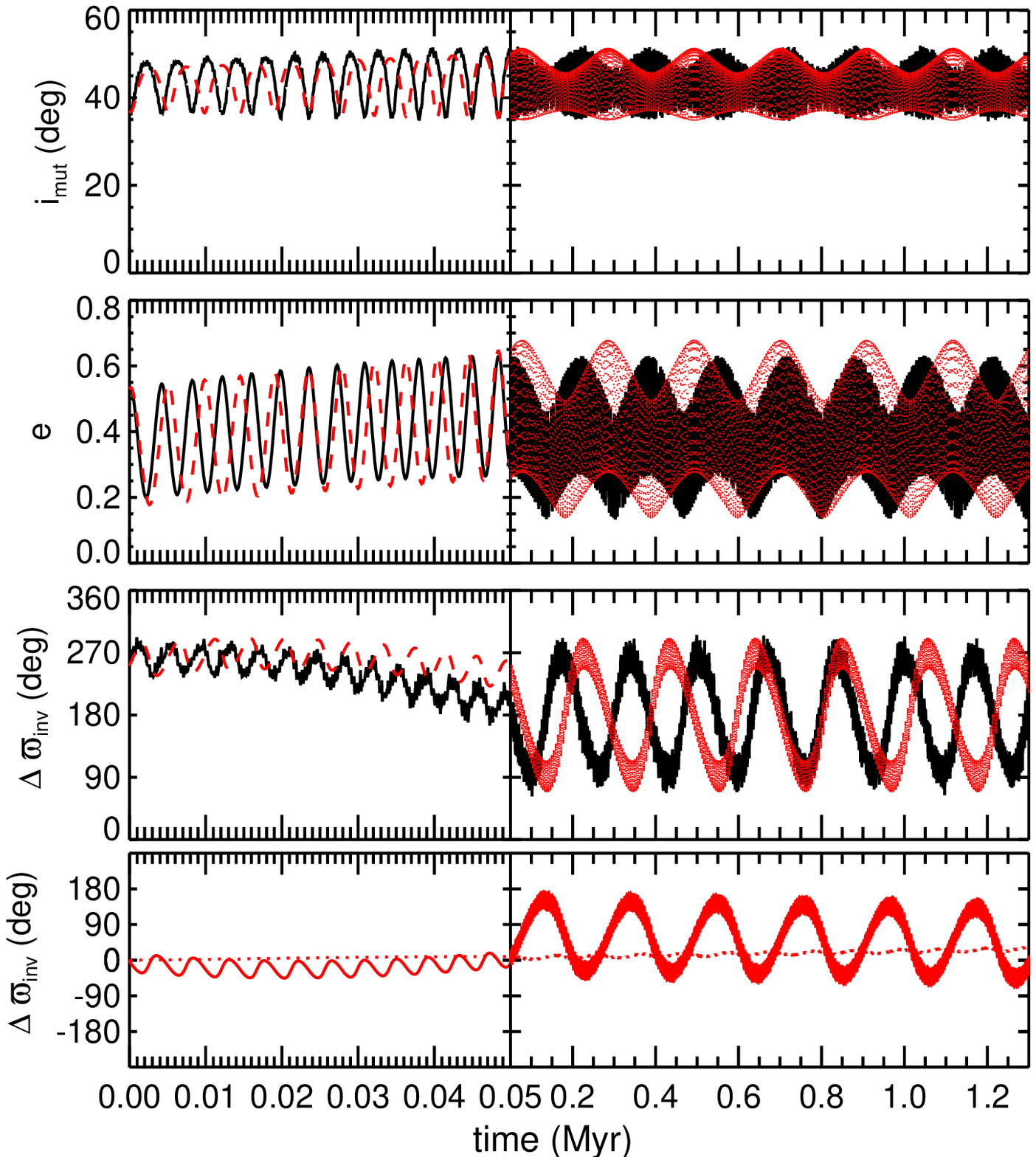}\\
{\noindent {\bf Fig.~S1.} \\
\noindent Orbital evolution of HD 168433b, calculated using an $N$-body simulation and using a secular Hamiltonian expanded to octupolar order.
Plotted are
  the time evolution of the mutual inclination (row 1), eccentricity
  of the inner warm Jupiter (row 2), apsidal separation (row 3), and
  near-canceling contributions to the apsidal separation from the
  quadrupolar and octupolar 
potentials
(row 4). The simulations were performed in the invariable plane. Initial conditions were for
  the HD 168433 system (Table S1) with  $i_{\rm sky} = 90^\circ$
and $\Delta \Omega_{\rm sky} = 37^\circ$, corresponding to $i_{\rm mut}$ = 37$^\circ$ and to angles
$\{i_1=33.1^\circ, \omega_1=263.0^\circ, \Omega_1=0^\circ\}$
and $\{i_2=3.9^\circ, \omega_2=334.3^\circ, \Omega_2=180^\circ\}$
in the invariable plane.
  All black
  curves in this figure are taken from the same {\tt Mercury6}
  $N$-body integration underlying the 
yellow curves in the left column of
  Fig.~2. Red curves are for a test particle integrated using
  Lagrange's equations of motion for the octupolar Hamiltonian
  \cite{yok03}. The bottom row was computed by separating the
  Hamiltonian into the quadrupolar terms (dotted line) and octupolar
  plus general-relativistic terms (solid line); by evaluating their
  respective contributions to $\dot{\varpi}_1$ using
  Lagrange's equations of motion; 
  and by integrating
  $\dot{\varpi}_1$, setting the constant of integration to zero
  for simplicity. (The quantities $e$ and $i_{\rm mut}$ were
  computed from the full Hamiltonian.) The sum of the dotted and solid
  curves in the bottom row equals the deviation of the red curve in the
  third row from its initial value. 
The secular behavior of the test particle 
reproduces qualitatively well that of the full $N$-body integration.}

\clearpage
\includegraphics[width=6in]{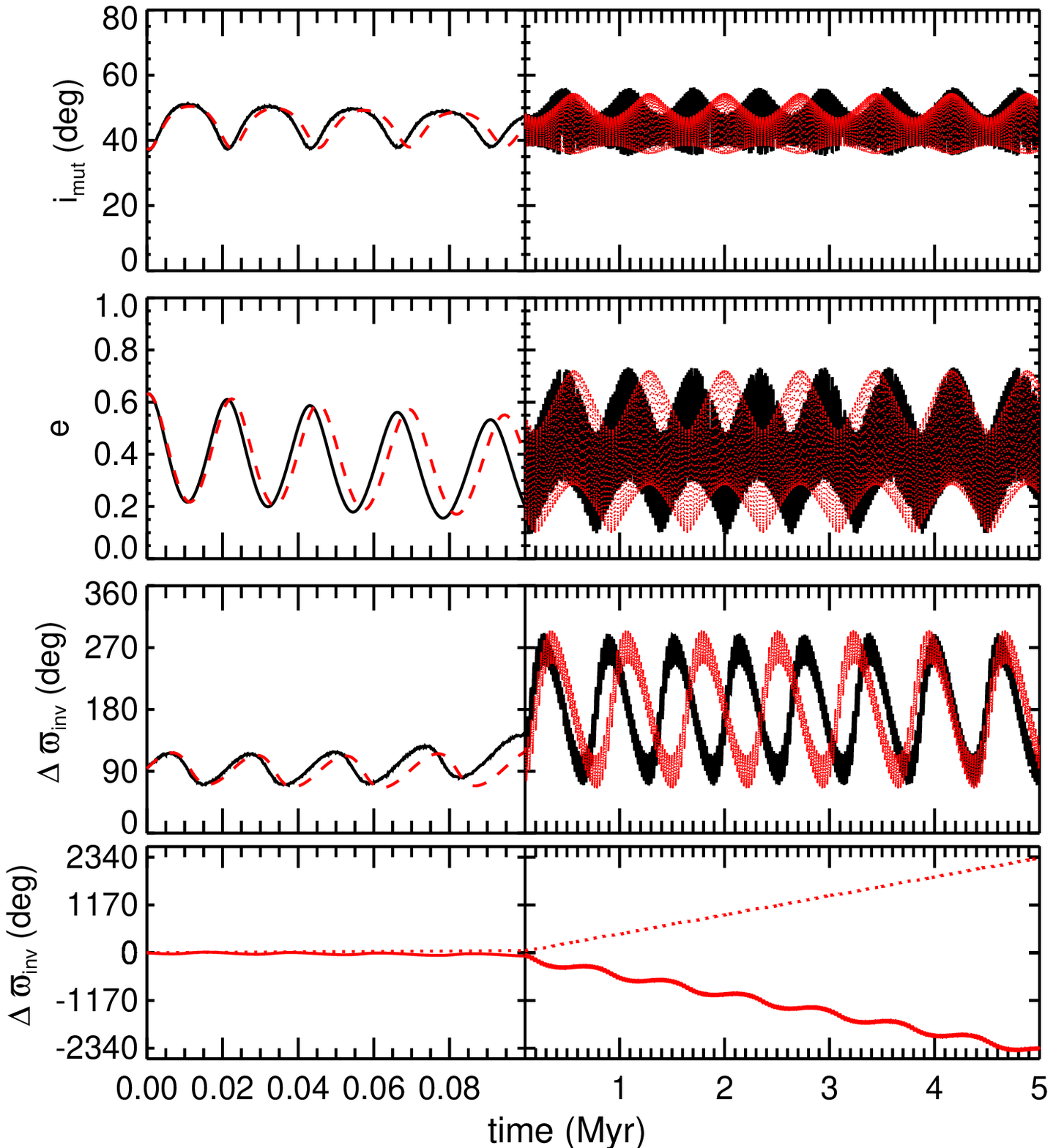}\\
{\noindent {\bf Fig.~S2.} \\
\noindent Orbital evolution of HD 74156b, calculated using an $N$-body simulation and using a secular Hamiltonian expanded to octupolar order.
Plotted are
  the time evolution of the mutual inclination (row 1), eccentricity
  of the inner warm Jupiter (row 2), apsidal separation (row 3), and
  near-canceling contributions to the apsidal separation from the
  quadrupolar and octupolar 
potentials
(row 4). The simulation was performed in the invariable plane. Initial conditions were for
  the HD 74156 system (Table S1) with $i_{\rm sky} = 90^\circ$
and
 $\Delta \Omega_{\rm sky} = 37^\circ$, corresponding to $i_{\rm mut}$ = 37$^\circ$ and to angles
$\{i_1=35.4^\circ, \omega_1=84.0^\circ, \Omega_1=0^\circ\}$
and $\{i_2=1.6^\circ, \omega_2=358.1^\circ, \Omega_2=180^\circ\}$
in the invariable plane. All black
  curves in this figure are taken from the same {\tt Mercury6}
  $N$-body integration underlying the
purple curves in the left column of
  Fig.~2. Red curves are for a test particle integrated using
  Lagrange's equations of motion for the octupolar Hamiltonian
  \cite{yok03}. The bottom row was computed by separating the
  Hamiltonian into the quadrupolar terms (dotted line) and octupolar
  plus general-relativistic terms (solid line); by evaluating their
  respective contributions to $\dot{\varpi}_1$ using
  Lagrange's equations of motion; 
  and by integrating
  $\dot{\varpi}_1$, setting the constant of integration to zero
  for simplicity. (The quantities $e$ and $i_{\rm mut}$ were
  computed from the full Hamiltonian.) The sum of the dotted and solid
  curves in the bottom row equals the deviation of the red curve in the
  third row from its initial value. 
The secular behavior of the test particle 
reproduces qualitatively well that of the full $N$-body integration.}
\clearpage

\clearpage
\includegraphics[width=6in]{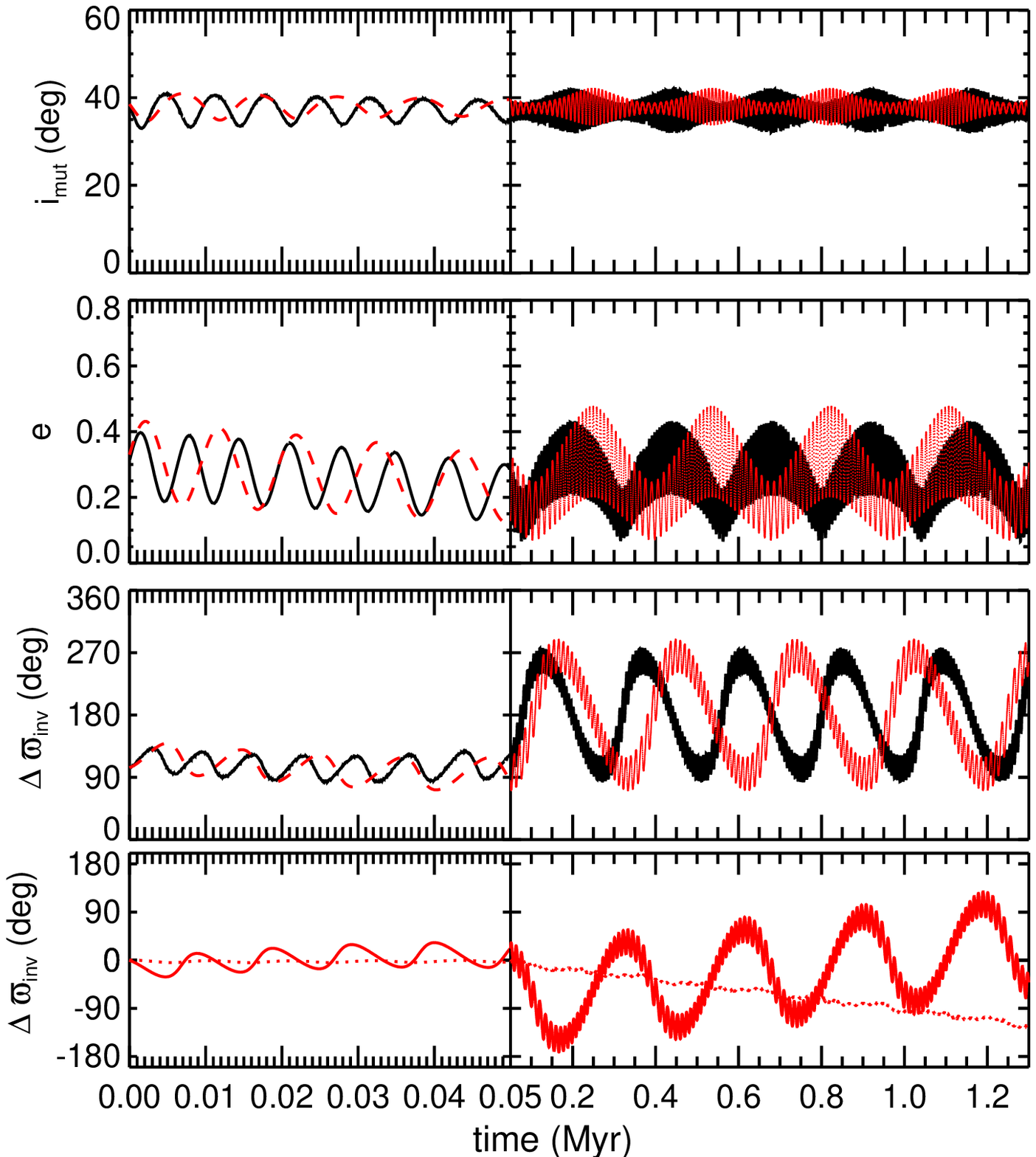}\\
{\noindent {\bf Fig.~S3.}\\
\noindent Orbital evolution of HD 169830b, calculated using an $N$-body simulation and using a secular Hamiltonian expanded to octupolar order.
Plotted are
  the time evolution of the mutual inclination (row 1), eccentricity
  of the inner warm Jupiter (row 2), apsidal separation (row 3), and
  near-canceling contributions to the apsidal separation from the
  quadrupolar and octupolar 
potentials
(row 4). The simulation was performed in the invariable plane. Initial conditions were for
  the HD 169830 system (Table S1) with $i_{\rm sky} = 90^\circ$
and $\Delta \Omega_{\rm sky} = 38^\circ$,
corresponding to $i_{\rm mut}$ = 38$^\circ$ and to angles
$\{i_1=28.6^\circ, \omega_1=238.0^\circ, \Omega_1=0^\circ\}$
and $\{i_2=9.4^\circ, \omega_2=162.0^\circ, \Omega_2=180^\circ\}$
in the invariable plane.
 All black
  curves in this figure are taken from the same {\tt Mercury6}
  $N$-body integration underlying the 
red curves in the left column of
  Fig.~2. Red curves here are for a test particle integrated using
  Lagrange's equations of motion for the octupolar Hamiltonian
  \cite{yok03}. The bottom row was computed by separating the
  Hamiltonian into the quadrupolar terms (dotted line) and octupolar
  plus general-relativistic terms (solid line); by evaluating their
  respective contributions to $\dot{\varpi}_1$ using
  Lagrange's equations of motion; 
  and by integrating
  $\dot{\varpi}_1$, setting the constant of integration to zero
  for simplicity. (The quantities $e$ and $i_{\rm mut}$ were
  computed from the full Hamiltonian.) The sum of the dotted and solid
  curves in the bottom row equals the deviation of the red curve in the
  third row from its initial value. 
The secular behavior of the test particle 
reproduces qualitatively well that of the full $N$-body integration.}
\clearpage

\clearpage
\includegraphics[width=6in]{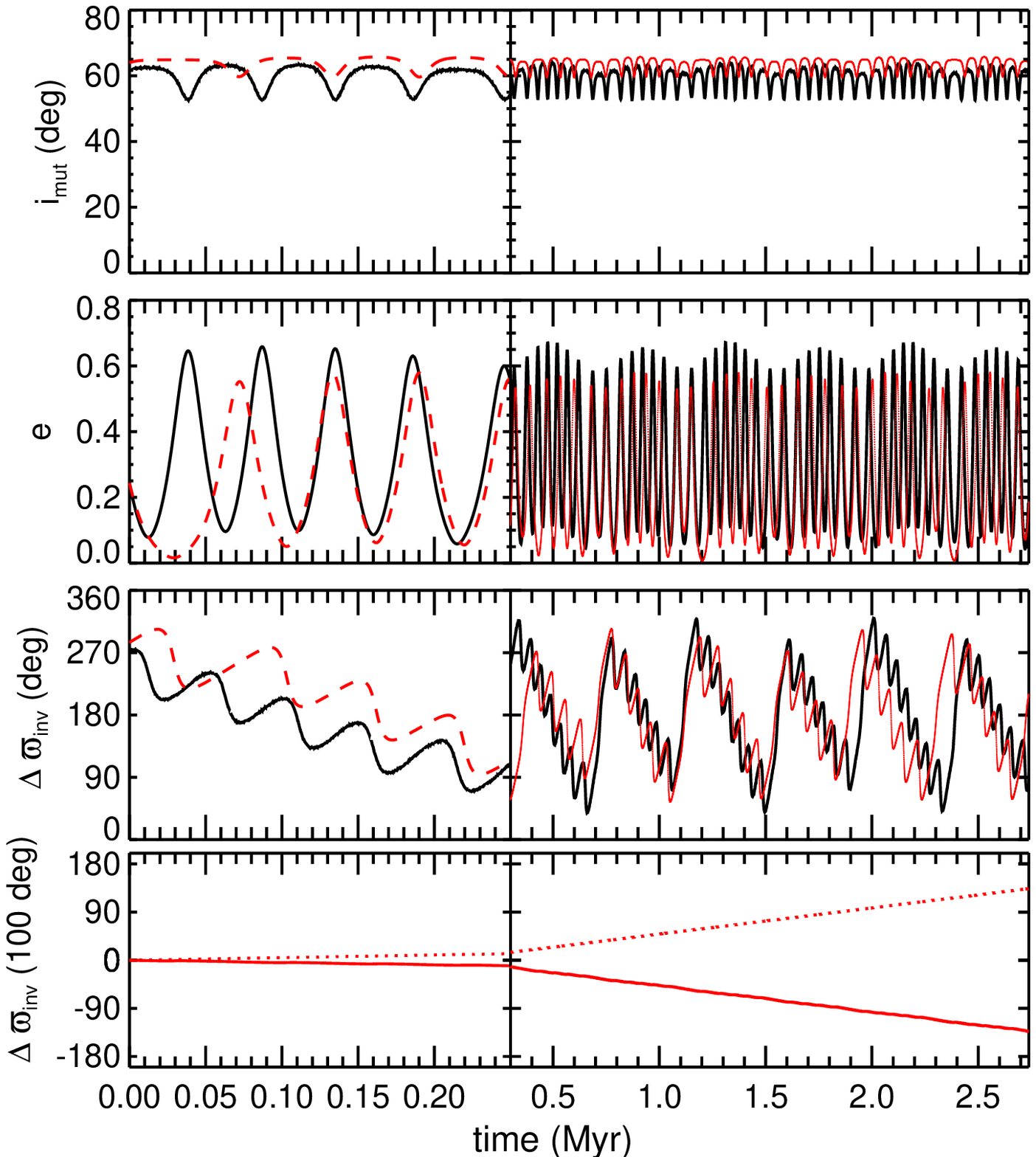}\\
{\noindent {\bf Fig.~S4.}\\
\noindent Orbital evolution of HD 38529b, calculated using an $N$-body simulation and using a secular Hamiltonian expanded to octupolar order.
Plotted are
  the time evolution of the mutual inclination (row 1), eccentricity
  of the inner warm Jupiter (row 2), apsidal separation (row 3), and
  near-canceling contributions to the apsidal separation from the
  quadrupolar and octupolar 
potentials
(row 4). Initial conditions were for
  the HD 38529 system (Table S1) with $i_{\rm sky, 1} = 43.7^\circ, i_{\rm sky,2} = 100.8^\circ, \Delta \Omega_{\rm sky} = 25^\circ$, corresponding to $i_{\rm mut}$ = 61.4$^\circ$ and to angles
$\{i_1=60.5^\circ, \omega_1=303.6^\circ, \Omega_1=0^\circ\}$ 
and $\{i_2=0.9^\circ, \omega_2=37.9^\circ, \Omega_2=180^\circ\}$ 
in the invariable plane. All black
  curves in this figure are taken from the same {\tt Mercury6}
  $N$-body integration underlying the
blue curves in the left column of
  Fig.~2. Red curves are for a test particle integrated using
  Lagrange's equations of motion for the octupolar Hamiltonian
  \cite{yok03}. The bottom row was computed by separating the
  Hamiltonian into the quadrupolar terms (dotted line) and octupolar
  plus general-relativistic terms (solid line); by evaluating their
  respective contributions to $\dot{\varpi}_1$ using
  Lagrange's equations of motion; 
  and by integrating
  $\dot{\varpi}_1$, setting the constant of integration to zero
  for simplicity. (The quantities $e$ and $i_{\rm mut}$ were
  computed from the full Hamiltonian.) The sum of the dotted and solid
  curves in the bottom row equals the deviation of the red curve in the
  third row from its initial value. 
The secular behavior of the test particle 
reproduces qualitatively well that of the full $N$-body integration.}

\clearpage
\includegraphics[width=6in]{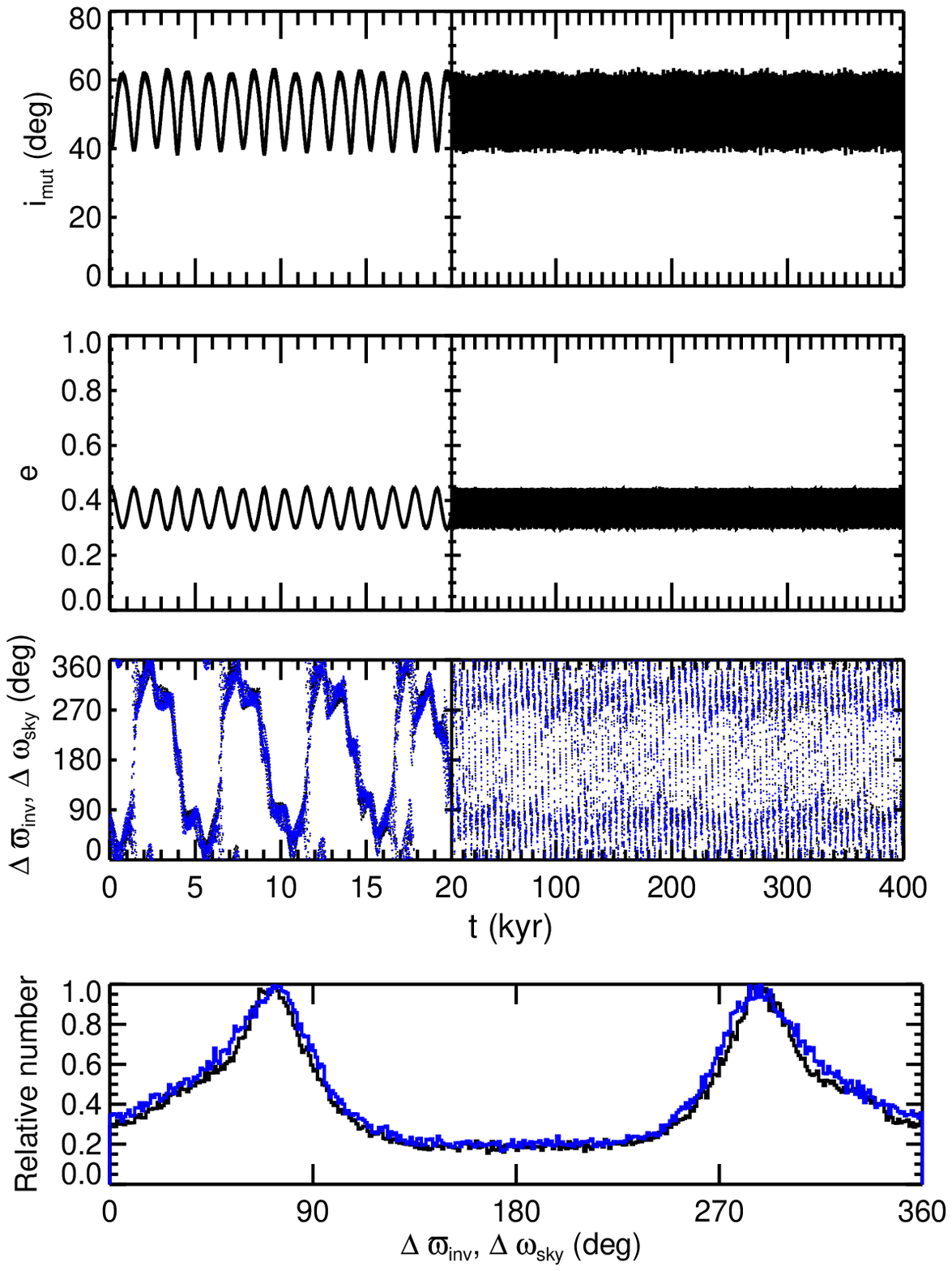}\\
{\noindent {\bf Fig.~S5.}\\
\noindent Orbital evolution of HD 202206, calculated using an $N$-body simulation performed in the invariable plane. The projection $\Delta \omega_{\rm sky}$ is overplotted in blue. Initial conditions were taken from the reported best-fit $N$-body solution\cite{cou10}, with the semi-major axis of the outer planet decreased (away from the 5:1 resonance) to the two-sigma lower limit from the observations,
$a_2 = 2.43$ AU
(see the second entry for HD 202206 in Table S1).
 The initial
orbital orientation
angles were $i_{\rm sky}=90^\circ$
and $\Delta \Omega_{\rm sky} = 41^\circ$ in the sky plane, corresponding to $i_{\rm mut} = 41^\circ$ and to $\{i_1 = 7.9^\circ, \omega_1=71.9^\circ, \Omega_1 = 311^\circ\}$ and $\{i_2 = 33.1^\circ, \omega_2=195.5^\circ, \Omega_2 = 131^\circ\}$
in the invariable plane. In contrast to the other five systems shown in Figs.~2--3 and Figs.~S1--S4, the HD 202206 system cannot be modeled in the test particle approximation using a purely secular Hamiltonian because of the system's proximity to the 5:1 resonance and the large relative mass of the inner planet 
($\min m_1= 16.59 M_{\rm Jupiter}$, $\min m_2=2.179 M_{\rm Jupiter}$).
Nevertheless, as illustrated in the bottom panel, the system still shows
a tendency to be found with its apsides nearly orthogonally misaligned.
}

\clearpage

\includegraphics[width=6in]{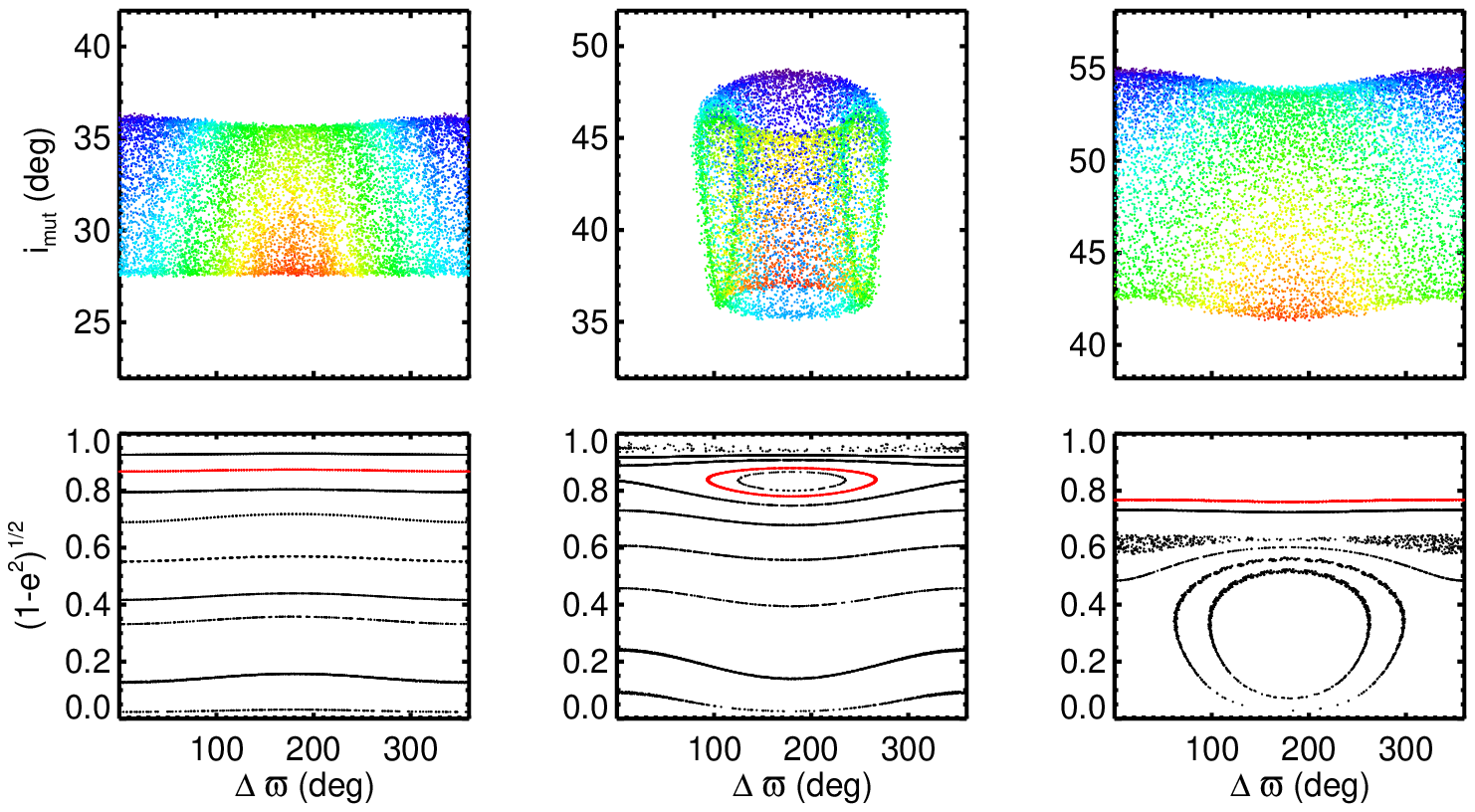}\\

{\noindent {\bf Fig.~S6.} \\
\noindent Trajectories of HD 147018b illustrating the lingering of
$\Delta \varpi_{\rm inv}$ near $90^\circ/270^\circ$ and libration of $\Delta \varpi_{\rm inv}$ about 180$^\circ$ for $i_{\rm mut} \sim 40^\circ$. 
Top row: Full trajectories from {\tt Mercury6} integrations performed in the invariable plane. The colors are contours of the inner planet's $z$-component of angular momentum. Each column represents an integration that starts from 
a different mutual inclination: $29^\circ, 39^\circ, 49^\circ$ from left to right (equivalently, $i_1 = 26.4^\circ, 35.6^\circ, 44.9^\circ$, respectively, in the invariable plane). All other initial conditions matched those in Fig.~3; the second column is the same integration as shown in Fig.~3,
and depicts the lingering of $\Delta \varpi_{\rm inv}$ near $90^\circ/270^\circ$
as two hollowed-out green lobes. We note that each column represents an individual trajectory; these are not Hamiltonian contour plots. Bottom row: Surfaces of section for the secular test-particle Hamiltonian \cite{yok03}, expanded to octupolar order and including general relativistic precession for the inner planet. Each panel corresponds to a value for the Hamiltonian equal to that in the trajectory in the row above, which is plotted in red; trajectories of equal Hamiltonian value but different initial conditions are plotted in black. Points are plotted whenever $\omega_1 = 90^\circ$. Only the middle panel features the sought-after libration of $\Delta \varpi$ about $180^\circ$ at the observed moderate $e_1$. The trajectories shown are selected by fixing $\varpi_2 = 316.865^\circ$.}

\clearpage
\includegraphics[width=6in]{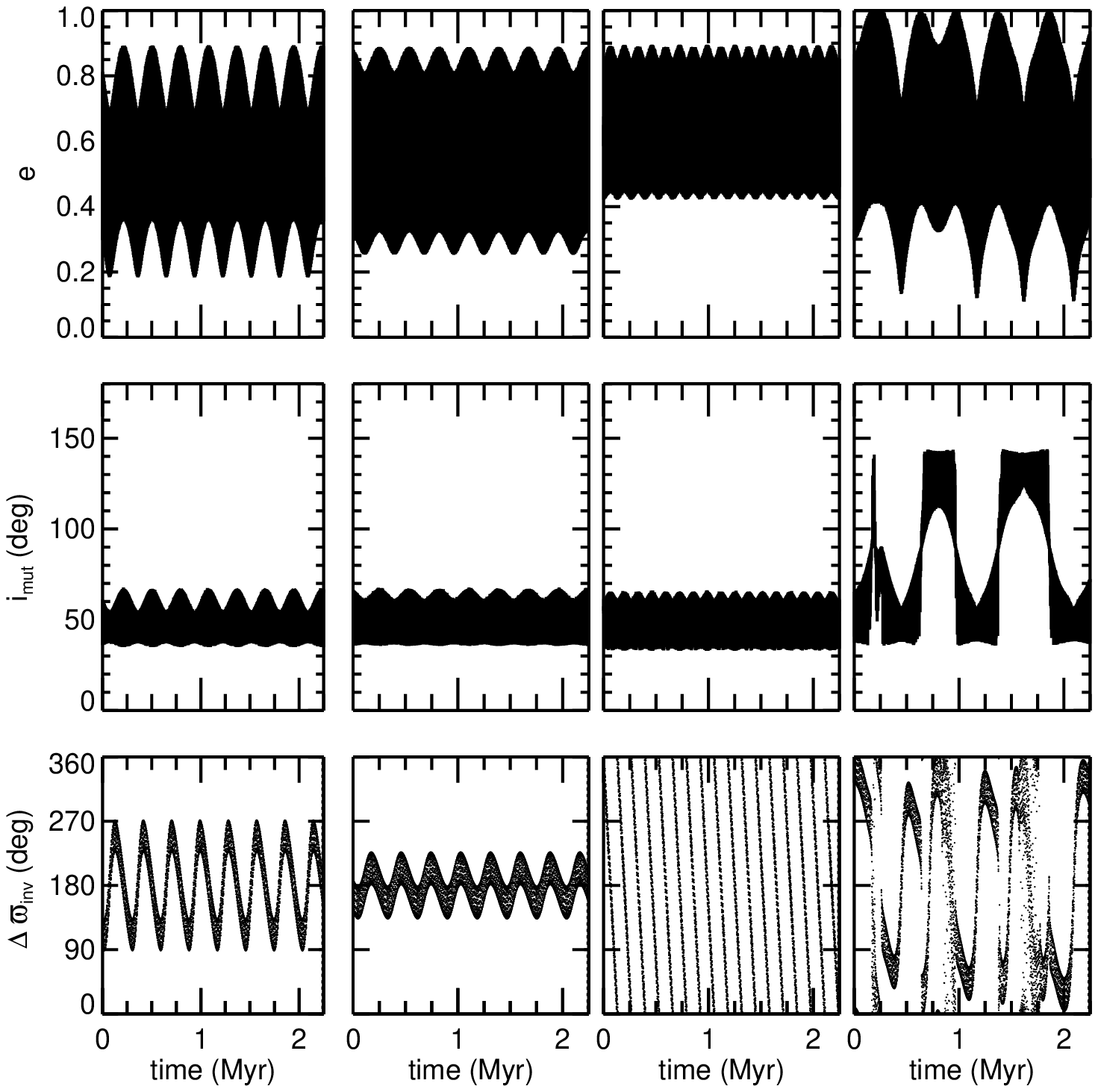}

{\noindent {\bf Fig.~S7.} \\
  \noindent Four integrations showing the range of behaviors possible
  for $\Delta \varpi_{\rm inv}$, $e$, and $i_{\rm mut}$. We used the
  test particle Hamiltonian expanded to hexadecapolar order
  \cite{yok03};
virtually identical results are obtained to octupolar order.
Initial
conditions of column 1 
were taken from $t= 150$ Myr of Fig.~4:
  $\{a_1 = 0.34 \,{\rm AU}, \Omega_1=160.3^\circ, a_2=1.923 \, {\rm AU}, e_2=0.133, i_2=0^\circ , \Omega_2 = 340.3^\circ\}$.
  Initial conditions for the other columns are the same except with $\{e_1=0.325, i_1=60.5^\circ, \omega_1=162.6^\circ, \varpi_2 = 90.1^\circ\}$ (column 1);
  $\{e_1=0.35, i_1=66^\circ,  \omega_1=162.6^\circ, \varpi_2 = 340.1^\circ\}$ (column 2);
  $\{e_1=0.65, i_1=55^\circ,  \omega_1=143.4^\circ, \varpi_2 = 320.9^\circ\}$ (column 3);  
  and   $\{e_1=0.325, i_1=65^\circ,  \omega_1=162.6^\circ, \varpi_2 = 322.9^\circ\}$ (column 4).  
The angle 
  $\varpi =\Omega- \omega $ when $i>90^\circ$ \cite{whi93}. Column 1:
  $\Delta \varpi_{\rm inv}$ lingers near 90$^\circ$/270$^\circ$ and
  the eccentricity has a strongly modulated envelope; the planet
  spends little time near its minimum periapse
(maximum $e$). 
Column 2: A smaller
  libration amplitude for $\Delta \varpi_{\rm inv}$ is accompanied by
  a less peaky modulation of eccentricity; the planet spends more time near
  minimum periapse as compared to Column 1.  Column 3: $\Delta
  \varpi_{\rm inv}$ circulates and eccentricity spends more time near its maximum value than in Column 1.
Column 4: For a large libration
  amplitude
for $\Delta \varpi_{\rm inv}$
and just slightly larger initial $i_{\rm mut}$, the planet
  is subject to inclination flips and eccentricity surges. When
  combined with tidal friction, the behavior in Columns 2--4 can lead
  to rapid tidal migration and 
the formation of 
hot Jupiters, whereas the behavior in
  Column 1 can produce stalled warm Jupiters.}

\clearpage
\begin{table}
\caption{Observed orbital elements
of seven RV-detected systems \cite{wri11} and one transit-detected system \cite{daw14} comprising warm ($0.1 < a \, ({\rm AU}) < 1$) Jupiters and their close friends ($a > 1$ AU). Only systems with a single pair of planets and 
securely measured
eccentricities and $\omega_{\rm sky}$ are listed \cite{select,unc}.}
\begin{tabular}{lllllllllll}
\\
System & $a_1$  & $a_2$ & $e_1$ &$e_2$ & $m_1$\ & $m_2$&$\omega_{\rm sky,1}$&$\omega_{\rm sky, 2}  $ &$|\Delta \omega_{\rm sky}|$& Ref.\\
&&&&& $\sin i_{\rm sky, 1}$& $\sin i_{\rm sky, 2}$\\\
&(AU)&(AU)&&&$ (M_{\rm Jupiter})$&$ (M_{\rm Jupiter})$&$(\circ)$&$(\circ)$&$(\circ)$\\
\hline
\hline
HD&\\
147018&0.239&1.92&0.469&0.133&2.127&6.59&336.0&226.9&109.1&\cite{seg10}\\
38529 &0.127&3.60&0.244&0.355&0.803&12.26&95.4&17.9&77.5&\cite{wri09}\\
168443\cite{corrected} &0.294&2.85&0.529&0.211&7.70&17.39&172.9&64.87&108.1&\cite{pil11}\\
74156 &0.292&3.90&0.630&0.380&1.773&8.25&174.0&268.0&266.0&\cite{mes11}\\
169830 &0.813&3.60&0.310&0.330&2.89&4.06&148.0&252.0&104.0&\cite{may04}\\
202206\cite{orig}&0.812&2.49&0.435&0.267&16.82&2.33&161.2&79.0&92.2&\cite{cor05}\\
202206\cite{best}&0.805&2.43&0.431&0.104&16.59&2.179&161.9&105.6&56.3&\cite{cou10}\\
82943\cite{third}&0.742&1.19&0.425&0.203&1.59&1.589&133&107&26&\cite{tan13}\\
\hline
Kepler-419&0.370&1.68&0.833&0.184&2.5&7.3&95.2&275.3&179.8&\cite{daw14}
\end{tabular}

\end{table}

\end{document}